\documentclass[letter,11pt]{article}
\pdfoutput=1
\usepackage{jinstpub} 
\usepackage{comment}
\usepackage{xcolor}

\title{Examining LEGEND-1000 cosmogenic neutron backgrounds in Geant4 and MCNP\newline}

\author[a,1]{C. J. Barton\note{Corresponding author.}}
\author[a]{W. Xu,}
\author[b]{R. Massarczyk}
\author[b]{S. R. Elliott}

\affiliation[a]{Department of Physics, University of South Dakota,\\Vermillion, SD, 57069}
\affiliation[b]{Los Alamos National Laboratory,\\Los Alamos, NM, 87545}

\emailAdd{cj.barton@roma3.infn.it}

\abstract{For next-generation neutrinoless double beta decay experiments, extremely low backgrounds are necessary. An understanding of in-situ cosmogenic backgrounds is critical to the design effort. In-situ cosmogenic backgrounds impose a depth requirement and especially impact the choice of host laboratory. Often, simulations are used to understand background effects, and these simulations can have large uncertainties. One way to characterize the systematic uncertainties is to compare unalike simulation programs. In this paper, a suite of neutron simulations with identical geometries and starting parameters have been performed with Geant4 and MCNP, using geometries relevant to the LEGEND-1000 experiment.
This study is an important step in gauging the uncertainties of simulations-based estimates. To reduce project risks associated with simulation uncertainties, a novel alternative shield of methane-doped liquid argon is considered in this paper for LEGEND-1000, which could achieve large background reduction without requiring significant modification to the baseline design.}


\newcommand{\novbb}{\ensuremath{0\nu\beta\beta}}
\newcommand{\gr}{$^{77}$Ge}
\newcommand{\gebeta}{$^{76}$Ge}

\begin{document}
\maketitle
\flushbottom

\newpage

\section{Motivation}
Observing the hypothetical neutrinoless double beta decay (\novbb) is the only experimentally feasible method to establish if neutrinos are Majorana particles, \textit{i.e.} fermions that are their own antiparticles. A positive observation has the potential to explain the origin of the beyond-the-Standard-Model neutrino mass and would violate total lepton number conservation ~\cite{Adams2022}. Tonne scale experiments with large exposure and extremely low backgrounds are required to probe the entire inverted mass ordering. Multiple tonne scale projects are in various stages of development using a wide range of decay isotopes and experimental techniques, including the Large Enriched Germanium Experiment for Neutrinoless $\beta\beta$ Decay (LEGEND)~\cite{Abgrall2021}, which searches for \novbb~in \gebeta~atoms. The LEGEND project has two phases. The initial phase, LEGEND-200, which will soon have 200 kg of active germanium detectors enriched in \gebeta, is operating in an early configuration at the Laboratori Nazionali del Gran Sasso (LNGS) in Italy~\cite{Baudis2022a}. The following phase of LEGEND, LEGEND-1000, will use 1000 kg of germanium detectors at a yet-to-be-determined host site to achieve a sensitivity to \novbb~ beyond a half-life of $10^{28}$ years~\cite{Abgrall2021}.

For most tonne scale \novbb~experiments, as well as other low-background rare event search experiments, neutron capture can lead to backgrounds which are difficult to veto. This is mainly due to the production of long-lived unstable isotopes, which can decay at some later time and be uncorrelated with other signals. As a result, these experiments need to carefully select detector materials to avoid radiogenic contributions~\cite{Arnquist2022a} and are hosted deep underground to reduce the secondary fast neutron flux from muons~\cite{Baudis2022}. Much attention has been paid by the community to these in-situ cosmogenic interactions at deep underground sites, including detailed simulation studies often carried out in Geant4. For instance, Geant4-based studies of LEGEND-1000 in-situ cosmogenic backgrounds have been reported in the preconceptual design report (pCDR)~\cite{Abgrall2021}. Notably, the liquid argon shield is both an extremely effective active veto~\cite{Agostini2022} and the main source of muon-induced neutrons, given that neutrons generated outside of the liquid argon shield are almost entirely moderated and absorbed in the large water shield. The ground and metastable states of \gr, which are produced by neutron captures on \gebeta, are determined to be the primary contributors to in-situ cosmogenic backgrounds for LEGEND after analysis cuts.

In-situ cosmogenic interactions, especially muon-hadron interactions, are complicated processes that are difficult to fully reproduce in controlled environments, and for which precise data is limited. Moreover, these interactions are very sensitive to the implementation of each experimental setup. Therefore, experimental results tend to be more informative for the design of similar experiments in the future. The LEGEND pCDR concludes that in-situ cosmogenic backgrounds could be a significant contribution for LEGEND-1000 at host sites with less overburden, within the large uncertainties that exist for these estimations. Analyses of data from the GERDA~\cite{Agostini2020} and \textsc{Majorana Demonstrator}~\cite{Arnquist2022ac} experiments have attempted to test Geant4 predictions for LEGEND, but both experiments~\cite{Vanhoefer2018,Arnquist2022b} have low counting statistics for the events of interest and only upper limits could be set. These experimental upper limits on the rate of cosmogenic \gr~production are roughly one order of magnitude higher than the predicted rate for LEGEND-1000, and the upper limits on $^{77m}$Ge production are significantly worse for both experimental data sets. Prioritized analysis of LEGEND-200 data could produce useful constraints once the experiment has accumulated sufficient exposure in a few years. At present, a cross-software comparison is critically important to increase confidence in simulation predictions, in particular for the predictions of neutron transport, interaction, and capture. These types of studies have been successfully performed before \cite{Ge2022,Affonso2020,Green2021}, but the comparisons are highly application-specific and a rigorous treatment will be tailored to the question at hand. 
It is also highly beneficial to envision alternative designs that can significantly eliminate in-situ cosmogenic backgrounds without requiring large modifications to the baseline design. The existence of alternative designs reduce project risk and could be useful for other rare event experiments.

The paper is organized in the following way. In Sec.~\ref{sec:g4_mcnp}, the two programs used in this study, Geant4 and MCNP, will be introduced, followed in Sec.~\ref{sec:simple} by comparisons using simple geometries, which will lay some basic expectations for both programs. In Sec.~\ref{sec:legend}, a representative geometry for LEGEND-1000 is implemented within both programs. Studies of in-situ cosmogenic backgrounds initialized in Geant4, but with neutron processes simulated in both programs, will be presented. Discussions of the comparison will be given in Sec.~\ref{sec:discuss}. In Sec.~\ref{sec:ch4}, an alternative design for LEGEND-1000 with methane doped Liquid Argon (LAr) will be proposed.

\section{Geant4 and MCNP}~\label{sec:g4_mcnp}

Geant4~\cite{agostinelli2003} is based on the C++ programming language, and incorporates an object-oriented approach to creating classes of particles, processes, and geometric objects. It has functionality to interface directly with the ROOT software distributed by CERN, providing a powerful built-in set of tools for recording and analyzing data. The wealth of information which can be retrieved about individual particles can result in intense memory and processor usage when executing the code, especially in very high energy applications such as in-situ cosmogenic simulations, so caution must be exercised when choosing which data should be recorded and which processes should be simulated in detail. \textsc{Majorana}, GERDA, and now LEGEND have used MaGe~\cite{boswell2011}, a Geant4 module tailored to suit the needs of these experiments and developed continuously for over a decade. There are other modules maintained by independent groups within the collaborations, such as the one used for this work, which uses Geant4 version 10.02.p02.

The Monte Carlo N-Particle (MCNP) radiation transport code is a proprietary software maintained and continuously developed for over 40 years by the X Computational Physics (XCP) group at Los Alamos National Laboratory (LANL)~\cite{Goorley2013}. Although it has a range of potential uses, MCNP is most widely known for applications in neutron-based physics, and especially reactor physics and criticality experiments. 
MCNP is rigorously benchmarked and tested internally by the XCP group before each new version release. However, the same principles must be considered for both MCNP and Geant4: every use case is unique, and for materials or situations which are not often encountered in the existing communities, exceptional care must be taken to ensure accurate results.

Although there have been many changes over its extensive lifetime, the current base languages for MCNP are Fortran 90 and C. Unlike Geant4, MCNP is an imperative programming language, and the principles behind creating and executing problems, as they are known in MCNP, are dissimilar from performing runs in Geant4. The typical output format is a text file containing a collection of data tables, which can be parsed by an analysis tool specified by the user, or manipulated to a limited degree by the built-in plotting subroutines for visualization. MCNP is designed to score tallied results and lacks the functionality to store event-by-event information for most parameters, reducing the ability to correlate information which can be retrieved about a single event, but also greatly reducing the processing and memory requirements of the program. This work uses release version 6.2 of MCNP.

Despite the differences in architecture and implementation, both Geant4 and MCNP adhere to the core tenets of Monte Carlo simulations. Individual events are propagated using random sampling to a significant degree, and using the central limit theorem, an average quantity or measurement can be obtained which could be interpreted as an estimated result for a similar physical system. Verifying the accuracy of the obtained result is the responsibility of the user, but properly managed, these programs provide powerful means to make predictions for currently unrealized projects.

In the Geant4 module, the 'shielding' pre-built physics list has been used to manage particle interactions and choose the appropriate data for determining cross-sections. For energies below 20~MeV, neutron processes are handled by the tabulated data available in G4NDL (neutron data library). For Geant4 versions 10.1 to 10.5, G4NDL4.5 is the default library, which bases its neutron cross-section data off of the ENDF/B-VII.1 (Evaluated Neutron Data Files) database~\cite{Chadwick2011}. MCNP chooses default neutron databases on a material-by-material basis using an internal system for determination, but the user may also specify the database to be used for each material. The MCNP materials in this study were explicitly assigned the ENDF/B-VII.1 database as well. The ENDF database is validated by a number of partner institutions in the Cross Section Evaluation Working Group, but its accuracy for any particular isotope is not guaranteed. However, the purpose of this study is not to validate the cross-section database employed, but rather to compare the propagation of particles in the programs under consideration, and any discrepancies in the database should be identical for both.

\section{Benchmarking}~\label{sec:simple}

The primary focus of this work is to compare neutron physics between Geant4 and MCNP. This information can be used to make general statements about the validity of the neutron simulations performed for the LEGEND experiment to date. When feasible, it is prudent to choose a simulation setup which is maximally sensitive to one aspect of neutron propagation, to isolate the potential differences between the two programs. Towards this end, simulations in a simple thin target geometry have been performed and evaluated.

To account for the wide array of energies at which muon-induced neutrons are produced, monoenergetic neutrons are simulated with initial energies spread across many orders of magnitude. For a LEGEND-1000-like geometry, the germanium detector array and the liquid argon shielding make up over 99\% of the total mass inside the cryostat. For that reason, the benchmarks were performed for these two materials, separately.

In nuclear and particle physics, fixed-target experiments have been standard practice for studying scattering effects for decades. A particle beam, often with a well-defined energy spectrum, is fired at a target, most commonly perpendicular to the surface normal, although beams with broad spectra and targets with angled surfaces are often studied as well\cite{Ramirez2017}. Detectors sensitive to the scattered incident particles or products of the scattering are used to recover as much information about the scattering as possible. Targets are typically limited to a few specific isotopes, such as the neutron-germanium cross-section studies in \cite{Bhike2015}. In a similar simulation setup, a thin target of either germanium or argon is bombarded with a monoenergetic beam of neutrons. To recover information about scattered particles, artificial bounding planes were placed enclosing the target, and the number of neutrons which are forward-scattered, back-scattered, or not interacting with the target are tallied. For scattered neutrons, the energy lost upon scattering can also be tallied.

\begin{figure}[h]
\centering
  \includegraphics[width=0.98\textwidth]{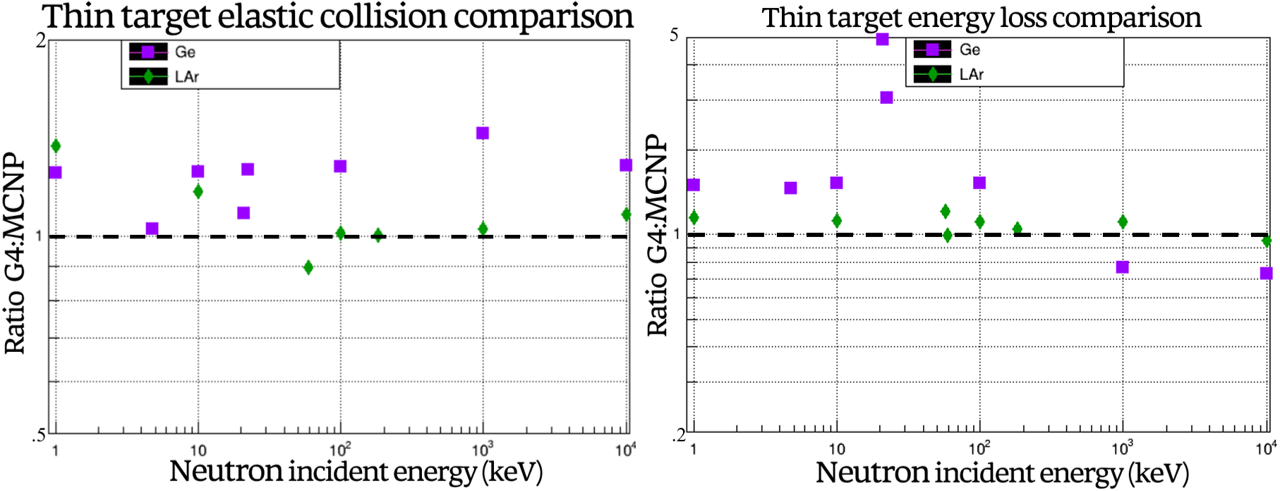}
  \caption{Ratio of results for average number of collisions (left) and average energy lost per collision (right) for a high-statistics thin target simulation in each material considered. A ratio greater than 1 means a higher rate in Geant4, and vice versa for a ratio lower than 1. The dotted line at ratio=1 indicates an exact match between the programs. For all visible data points, statistical error is below 5\%, and is generally less than 1\%.}
  \label{fig:thintargetresults}
\end{figure}

Figure \ref{fig:thintargetresults} summarizes the results of the thin target simulations. The scattering plots are presented as a dimensionless ratio of the results from Geant4 and MCNP, to facilitate comparison.
For both materials, neutrons were initialized with starting energies increasing by one order of magnitude, from 1 keV to 10 MeV, which appear along the vertical grid lines. Data points not aligned to the vertical grid lines are known neutron scattering resonances or anti-resonances in the chosen ENDF database for the respective materials.

\begin{figure}[h]
\centering
  \includegraphics[width = 0.8\textwidth]{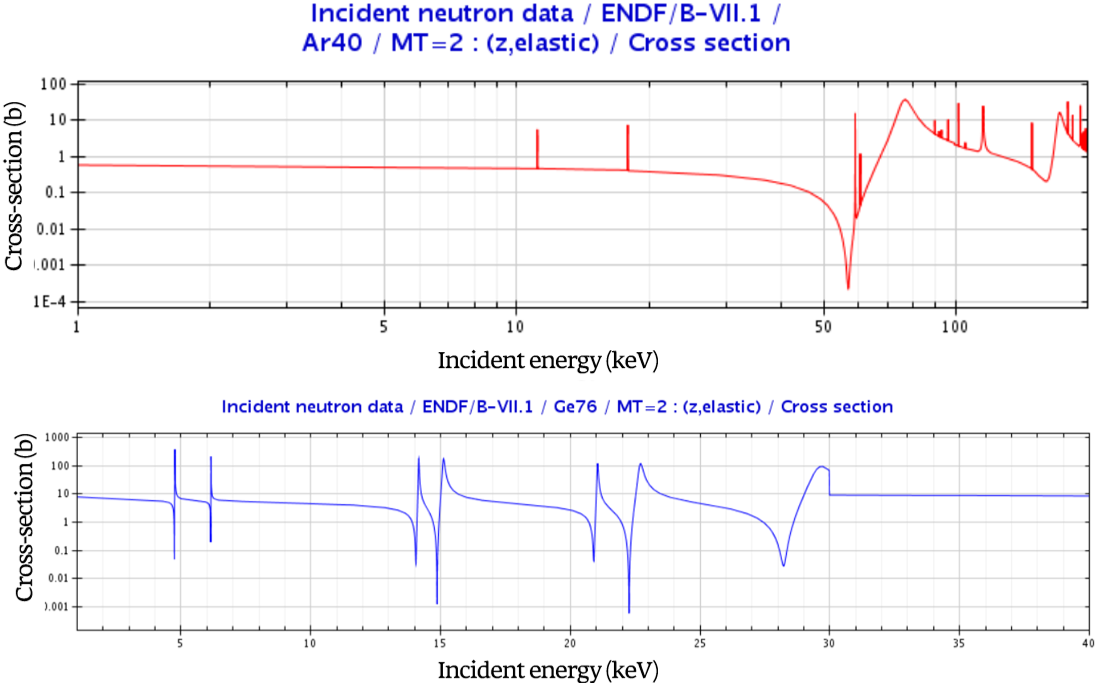}
  \caption{Elastic scattering cross-sections and (anti)resonances for $^{40}$Ar (top) and $^{76}$Ge (bottom). The germanium detectors contain 90\% of this isotope and 10\% $^{74}$Ge. (Anti)resonance lines can vary between existing neutron databases. Data plotted is from the ENDF/B-VII.1 neutron database used in this study.}
  \label{fig:elasticxcs}
\end{figure}

 Results indicate a higher average scattering cross-section in Geant4 for all starting energies in both materials, except for the 59.3 keV resonance in liquid argon. There is a scattering anti-resonance for liquid argon at 57.1 keV, recently confirmed by the ARTIE experiment at LANL as a sharp dip in the cross-section between 50 and 60keV~\cite{Andringa2022}. The ratio of Geant4 to MCNP cross-section at this energy was determined to be 41.7, and this outlier is not plotted. Although statistics were limited by the low cross-section in both cases, this difference is far outside what may be accounted for by statistical uncertainty (calculated to be about 35\%). As seen in Figure \ref{fig:elasticxcs} of cross-sections in popular databases, the behavior of the cross-section near this energy varies rapidly, and although both sets of simulations used the ENDF database, even small differences in how the scattering data is tabulated in this region may lead to dissimilar results for monoenergetic neutrons.

For the average neutron energy loss from each scattering event, Geant4 almost matches MCNP in the liquid argon, but trended substantially higher in the germanium target for most of the neutron incident energies considered. However, for higher incident energies, energy loss per scatter in germanium was greater in MCNP. Also, the resonance at 21.1 keV and the anti-resonance at 22.3 keV in germanium both lost much more energy per scatter in Geant4. In principle, there should be no connection between the tabulated scattering rate and the energy lost per scatter, but it is suggestive that of the 15 data points plotted, the only two major outliers are at a resonance and an anti-resonance in the same material.

Since Geant4 trends higher in both average scattering rate and average energy loss below 1 MeV, neutrons will most likely lose energy faster in this program when propagated through large volumes of these materials, such as would be found in the LEGEND-1000 cryostat. On the other hand, lower energy loss and higher scattering cross-section in Geant4 for neutrons in the MeV energy range complicate the discussion in this energy region, which is relevant for the $0\nu\beta\beta$ Q-values. Direct scattering of muon-induced neutrons can be effectively vetoed in the proposed LEGEND-1000 experimental setup. The main concern remains with delayed backgrounds due to neutron captures.

\section{LEGEND-like geometry}~\label{sec:legend}

\begin{figure}[h]
\centering
  \includegraphics[scale=0.4]{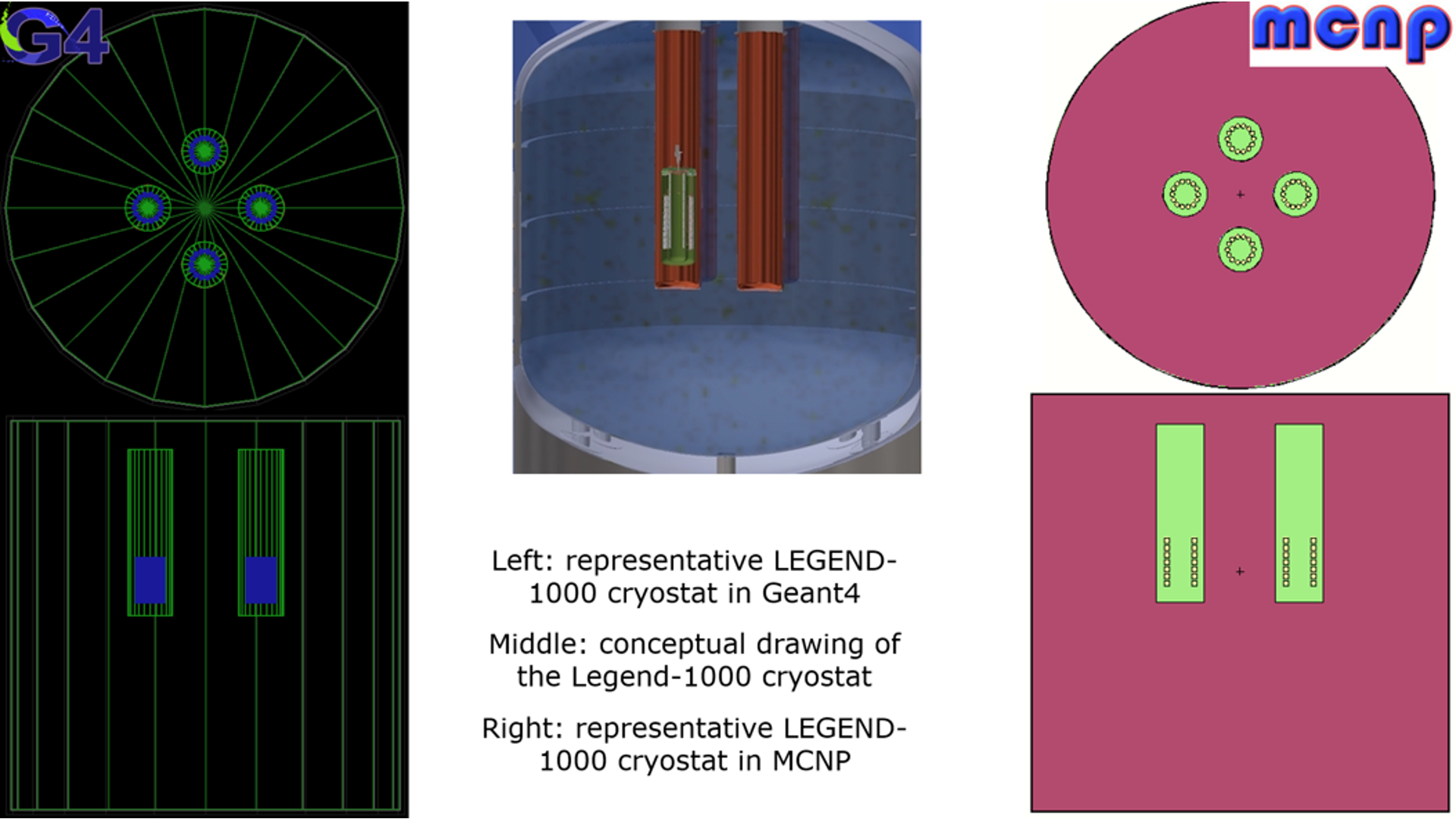}
  \caption{Cross-sections of the geometry in each program, as well as a reference design.}
  \label{fig:comparegeom}
\end{figure}

For an answer which is relevant in the context of LEGEND-1000, a LEGEND-1000-like geometry must be implemented. Towards this end, a simplified version of the pCDR baseline design cryostat has been created in both Geant4 and MCNP. While Geant4 once again takes an object-oriented approach to implementing geometries, with access to C++ functionality and class structure, MCNP uses a combinatorial geometry in which the user specifies first and second fundamental forms of curvature on a coordinate system and defines the relation of these objects in a logical manner. 
Figure \ref{fig:comparegeom} displays a cross-section of the geometry as it appears in each program, along with a reference drawing of the LEGEND-1000 baseline cryostat. A stainless steel outer tank encapsulates the bulk liquid argon shielding. The four reentrant tubes are copper cylinders which encapsulate the liquid argon closest to the detector arrays. All dimensions are made to the exact same specifications in each program's geometric implementation.

\subsection{Implementation, limitations and approximations}
The MUSUN muon simulation software\cite{Kudryavtsev2009} was used to generate a set of 25 million muons with energy and angular distributions similar to those at Gran Sasso National Laboratory (LNGS), the host site for LEGEND-200 and a candidate site for LEGEND-1000. These muons were propagated through the simplified cryostat in this work, using Geant4. When a free neutron was created, its energy and position were recorded, and the neutron was subsequently removed from the rest of the simulation. This is to prevent a double-counting issue which can occur when a neutron scatters inelastically.

The ROOT file containing the information recorded about the muon-induced neutrons is used to make a series of energy and position distributions. These distributions are necessary because, unlike Geant4, MCNP cannot read ROOT files to import information about individual particles on an event-by-event basis. To keep the comparison exact, Geant4 was therefore limited to reading neutron information from these distributions as well.

\begin{figure}
\centering
  \includegraphics[scale=0.7]{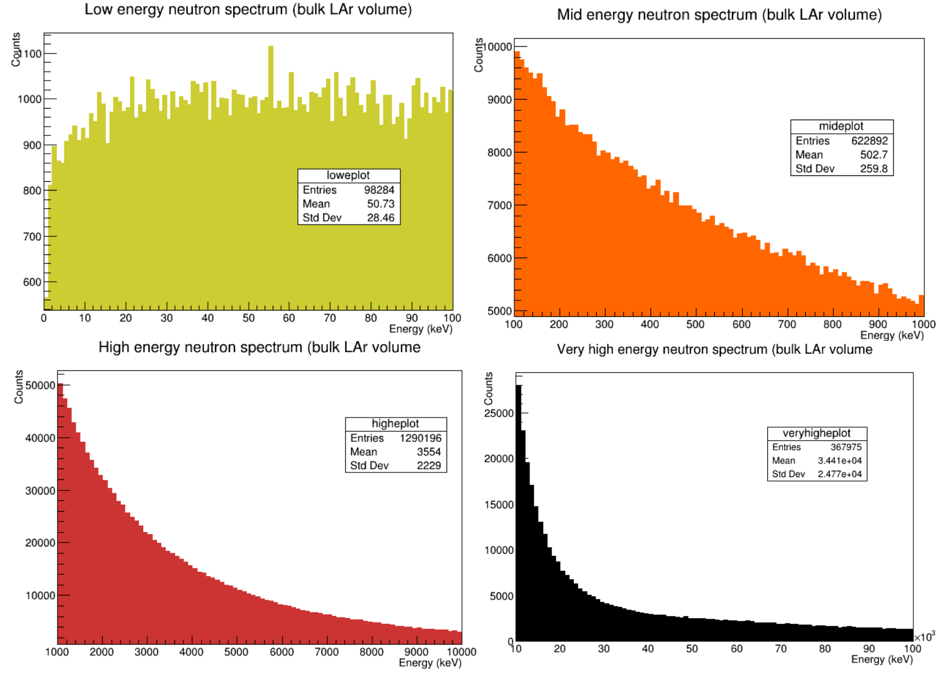}
  \caption{The energy spectrum of muon-induced neutrons. Since these neutrons span several orders of magnitude in initial kinetic energy, four simulations have been performed (0-100 keV, 100-1000 keV, 1-10 MeV, and 10-100 MeV). All plots are in keV, with the bin size of each distribution equal to 1\% of its maximum energy.}
  \label{fig:muonneutronespec}
\end{figure}

\begin{figure}
\centering
  \includegraphics[scale=0.7]{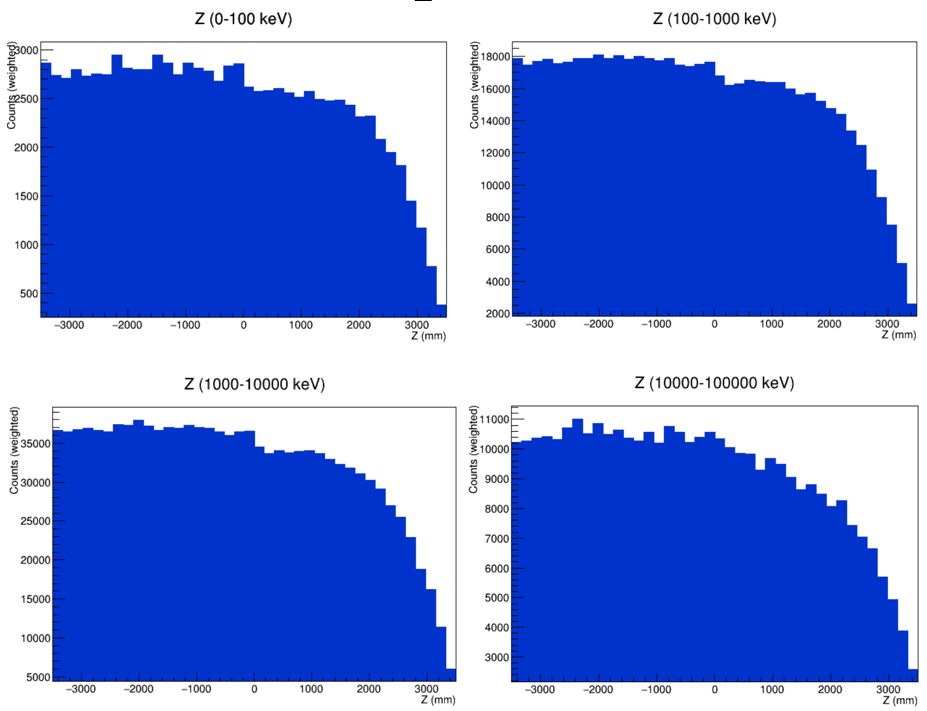}
  \caption{The distribution of starting z position for muon-induced neutrons. A distribution has been made for each energy range in Figure \ref{fig:muonneutronespec}. The step-like behavior near z=0 is due to shadowing from the copper tubes and detectors at positive z values.}
  \label{fig:muonneutronzspec}
\end{figure}

Four sets of energy distributions were made, as seen in Figure \ref{fig:muonneutronespec}. Each has a corresponding vertical position distribution (Figure \ref{fig:muonneutronzspec}). The four initial neutron energy ranges considered are from 0 to 100 keV ('low energy'), 100-1000 keV ('medium energy'), 1-10 MeV ('high energy'), and 10-100 MeV ('very high energy'). Neutrons are produced more often at lower positions in the tank, since the muons travel from the top downward, and the secondary particle showers tend to expand with increased path length in the argon. The radial distribution of the neutrons was assumed to be approximately flat per unit volume. This introduces a 10\% uncertainty, but reduces the challenges associated with implementing a radial distribution into MCNP, which is naturally Cartesian. In addition, MCNP cannot have an imported distribution which is dependent on another imported distribution, so attempting to characterize position in more than one dimension might prove unwieldy.

It should also be noted that the ENDF library used to determine neutron scatter and capture probabilities is limited to neutrons with kinetic energies under 20 MeV. For neutrons with more than 20 MeV kinetic energy, a parametrized model is implemented in Geant4, and MCNP does not provide interaction models above 20 MeV by default. Results from the 'very high' energy range of recorded neutrons, which spans from 10 MeV to 100 MeV, cannot be systematically compared but are included for completeness. The comparison of elastic scattering behavior remains robust for the 10-20 MeV neutrons, which comprise about half of the neutrons in this energy range (see Figure \ref{fig:muonneutronespec}), but other interaction channels which become relevant at energies above 10 MeV have not been included in this comparison study, so the results cannot be used. The purpose of this study is to evaluate the effectiveness of the simulation tools used to estimate muon-induced neutron backgrounds for LEGEND, which occur via neutron capture. Since capture occurs much more frequently for neutrons with low kinetic energies, these very high energy neutrons do not contribute directly to the total capture rate, and do not significantly impact the remainder of the study. It is more likely for neutrons of such energies to escape the cryostat entirely and be thermalized and captured in the water shield surrounding the cryostat, which has no effect on the background model of LEGEND-1000. On the other end of the energy spectrum, it is possible that the handling of neutrons near thermal energies is not identical. All materials under consideration have been set to a temperature of 87 Kelvin in both programs, but it is unclear if the statistical treatment of thermalized neutrons can be considered equivalent with the information which has been collected.

As a final note, when specifying the output tables given by MCNP, it is the responsibility of the user to provide the range, bin count, and type of parameter which is to be tallied. With Geant4 it is possible to 'redraw' the data while manipulating it with ROOT to increase or decrease the bin count, change an axis to logarithmic scaling, etc, but these choices must be made prior to execution in MCNP. Whenever possible, Geant4 output has been manipulated to conform to what was chosen in MCNP, to take advantage of the additional flexibility offered in the former.

\subsection{Results}
\begin{figure}[h!]
  \begin{minipage}[L]{0.45\textwidth}
\begin{centering}

  \includegraphics[width=0.99\textwidth]{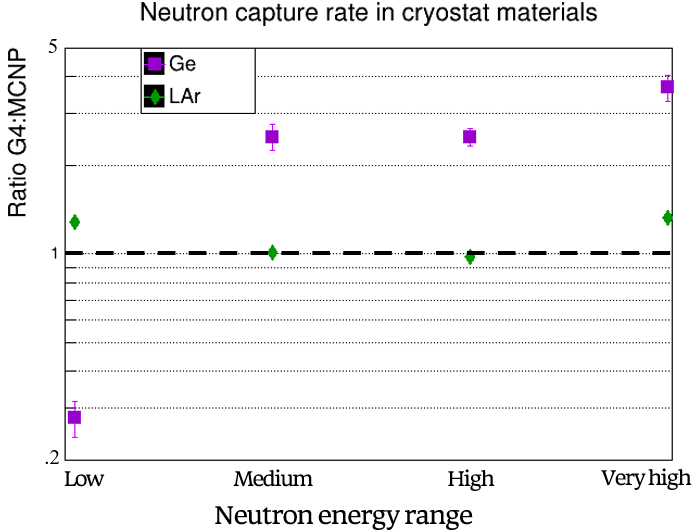}
  \caption{Ratio of neutron capture rates in germanium (Ge, purple squares) and liquid argon (LAr, green diamonds) for the two simulation programs. Only statistical errors are plotted.}
  \label{fig:muonneutroncaptures}

\end{centering}
  \end{minipage}
  \begin{minipage}[R]{0.45\textwidth}
  
    \includegraphics[scale=0.5]{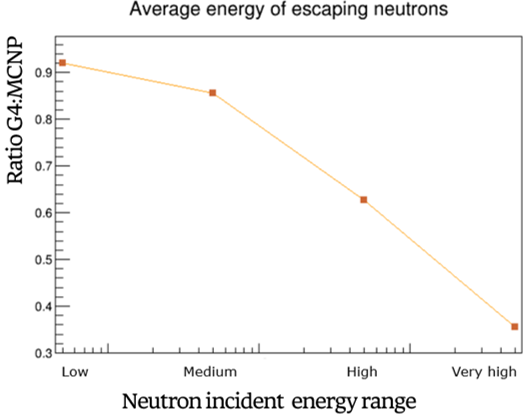}
  \caption{Ratio of average neutron energy of escaped neutrons. A ratio below 1 indicates that the average remaining energy is higher in MCNP than in Geant4.}
  \label{fig:muonneutronenergy}
  
  \end{minipage}
\end{figure}

\begin{figure}
\centering
  \includegraphics[scale=0.45]{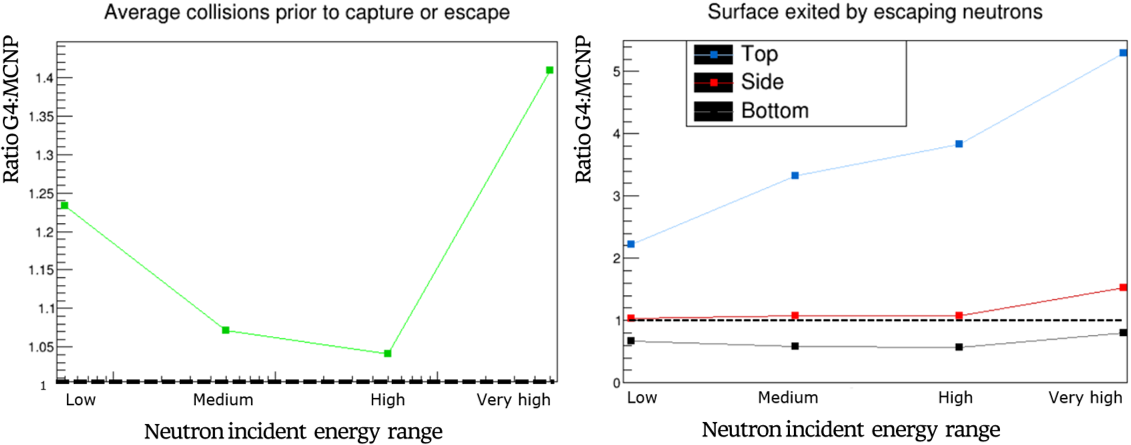}
  \caption{Comparison of average collisions experienced by all neutrons (left), and which outer surface escaping neutrons pass through (right). Once again presented as the ratio of the Geant4:MCNP results.}
  \label{fig:muonneutronescape}
\end{figure}

The results of this set are summarized in Figures \ref{fig:muonneutroncaptures}, \ref{fig:muonneutronenergy} and \ref{fig:muonneutronescape}. As with the thin target set, the plots are presented as a dimensionless ratio of each data point in the two programs. From previous trends, the expectation is for a higher scattering and capture rate in Geant4 than in MCNP for most energies, resulting in a ratio greater than 1. Each simulation represents ten years of muon-induced neutron data at LNGS depth, and error bars for Figure \ref{fig:muonneutroncaptures}~were calculated using whichever simulation program had lower statistics at each data point.

The most important parameter of interest in these comparisons is the neutron capture rate on the germanium detectors, \textit{i.e.} the production of \gr~states, as this quantity effectively determines the expected rate of neutron-induced background signals in the analysis region for \novbb. As seen in Figure \ref{fig:muonneutroncaptures}, the capture rate in germanium is higher for three of the neutron initial energy ranges, which represent over 95\% of the total muon-induced neutrons. The capture rate in the bulk liquid argon volume is either comparable or slightly higher in Geant4.

Since the neutron capture cross-section is energy dependent, it is intrinsically tied to the scattering rate and the energy lost per scatter as well. Figure \ref{fig:muonneutronescape} demonstrates that the average number of collisions is higher in Geant4 than in MCNP for all energy ranges, and Figure \ref{fig:muonneutronenergy} plots the average kinetic energy of an escaping neutron. The average energy upon escape is lower in Geant4 for all energy ranges, with the difference growing more pronounced at increased energies. 

Finally, it is worth noting that the cylindrical cryostat's outer surface is divided into three sections: the flat top and bottom surfaces, and the curved side surface. As in the right side of Figure \ref{fig:muonneutronescape}, the ratio of neutrons exiting from the top surface is dramatically higher in Geant4, with the disparity once again increasing for higher energy ranges. It is also worth noting that the top surface is the least likely exiting surface, due to the neutrons being produced mostly at lower positions in the argon tank.

\section{Discussion}~\label{sec:discuss}
As has been noted a throughout this report, neutrons tend to scatter more frequently in Geant4 for the materials under consideration. Additionally, the average energy loss per neutron scatter is higher in Geant4 in general. These effects compound, resulting in higher neutron capture rate and higher energy loss in the muon-induced neutron simulation for Geant4, compared to MCNP. This suggests that previous simulations work performed for LEGEND over the past few years using Geant4 version 10 will have more conservative results than would have been determined using MCNP. 

For the muon-induced neutron simulations, an inverse linear relationship would be expected between average number of scatters and average remaining energy of escaped neutrons, if the energy loss kinematics and capture cross-sections were the same in both programs. However, at increasing energies, Geant4 neutrons continue to escape with subsequently lower energies compared to MCNP. The capture cross-sections aren't dramatically different in the bulk argon material. Due to this and Geant4's tendency to scatter neutrons through the top of the cryostat with greater frequency, the current hypothesis is that neutrons are scattered 'harder' in Geant4, and tend to have wider average scattering angles than in MCNP. However, with the information currently available, this is a difficult hypothesis to verify.

There are a few outstanding discrepancies between the two programs which remain unresolved or poorly understood. With the information available, the two programs have been compared as faithfully as achievable, and some conclusions can be drawn or inferred about the similarities and differences in neutron behavior in Geant4 and MCNP. Any future work would utilize more advanced analysis techniques in MCNP to attempt to gather more information than is easily retrieved. Analysis of neutron captures forces 'analog' simulations in MCNP, disabling most of the more powerful variance reduction features available as well as some of the advanced pointwise analysis techniques. A separate identical simulation with neutron capture disabled would allow access to these additional techniques for further study.

The first phase of the LEGEND project, LEGEND-200, is currently operational at LNGS. With the data collected from LEGEND-200, it will be possible to characterize to some degree the deep underground muon-hadron interactions through the correlated argon scintillation and germanium solid-state detector data. This will eventually lead to estimates which can support or constrain the predictions in this work.

\section{Methane Doping in Liquid Argon}~\label{sec:ch4}
Studies presented here suggest that the Geant4 prediction of in-situ cosmogenic backgrounds in the LEGEND-1000  pCDR is more conservative than an MCNP prediction would have been. However, the most relevant experimental data will come from LEGEND-200 after a few to several years of data collection. Due to the large uncertainties present in cosmogenic simulations, the experimentally determined cosmogenic background rate could be significantly higher than simulation predictions, or could even be significantly lower. It is therefore prudent to devise alternatives for LEGEND-1000 which will suppress the cosmogenic background further, if necessary. Ideally, a minimally invasive modification could be implemented only if necessary, and thus the project risks can be significantly reduced without impacting the project progress. This paper proposes an option to heavily dope the outer liquid argon volume of LEGEND-1000 with hydrogen-rich materials. Hydrogen is a well-known neutron moderator, with a significant neutron scattering cross-section across a wide range of kinetic energies. Once the neutrons are sufficiently moderated, they can be absorbed in the liquid argon volume rather than in the germanium crystals, preventing the production of \gr~and thus reducing the background induced by cosmogenically-induced neutrons.

The LEGEND-1000 experiment envisions utilizing two types of liquid argon: atmospheric liquid argon (atmLAr) and underground liquid argon (UGLAr). The atmLAr is the typical liquid argon that is condensed from the atmosphere. It is readily available and widely used in many modern experiments. However, cosmic bombardments in the upper atmosphere produce $^{42}$Ar atoms, which will ultimately contribute to the LEGEND backgrounds via the progeny $^{42}$K~\cite{Abgrall2021}. UGLAr is extracted from deep underground reservoirs. With a half-life of only 33 years, $^{42}$Ar atoms in these reservoirs have largely decayed away, and UGLAr can provide a much more radiopure shielding material for LEGEND-1000. Due to the difficulty of extracting UGLAr, it is only used in the smaller liquid argon volumes immediately surrounding the germanium detector arrays, which are encapsulated by the copper reentrant tubes. The reentrant tubes separate UGLAr from atmLAr as shown in Figure~\ref{fig:comparegeom}. Only the UGLAr in the reentrant tubes is required to serve as an active veto in the nominal LEGEND-1000 design, leaving the much larger amount of atmLAr outside of the reentrant tubes as a passive shield without instrumentation for light collection. This is an important design feature, since hydrogen-rich dopants tend to quench scintillation light. The fact that LEGEND-1000 does not strictly require light collection for the majority of its LAr shielding is a unique opportunity for heavy doping with suitable materials.

Methane (CH$_4$) can be doped into liquid argon at fractions of up to 50\% by volume (44\% molar fraction)~\cite{Anderson1988} without losing solubility. The high solubility of methane has been confirmed for a LAr system operated under typical conditions, such as a saturated vapor pressure of 1.00 atm~\cite{Bondar2020, Bondar2022}. Therefore, it is feasible for LEGEND-1000 to heavily dope the atmLAr outside of the reentrant tubes, the so-called outer LAr volume. This will not require significant modifications to the cryostat design, and would leave the UGLAr active veto unchanged. 

\begin{figure}[t]
\centering
  \includegraphics[width=0.94\textwidth]{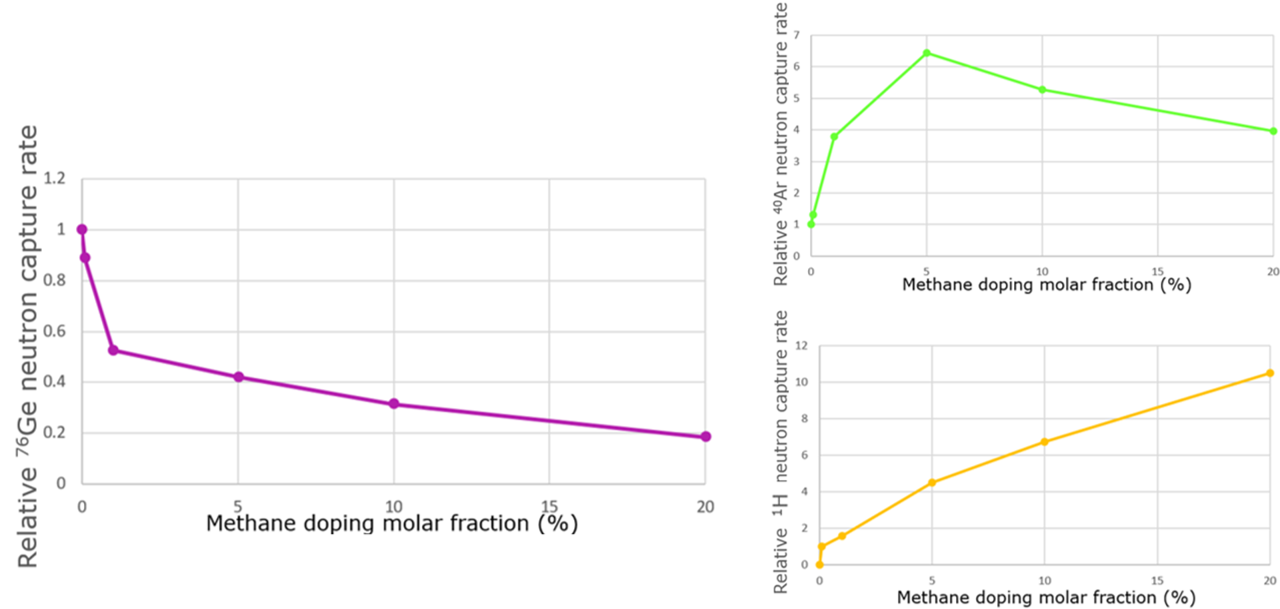}
  \caption{Neutron capture rate in three isotopes, as a function of molar fraction of methane doping. The capture rate for the $^{76}$Ge and $^{40}$Ar are normalized to the undoped case (0\%). There is no hydrogen in the undoped case, so the third plot is normalized to the lowest doping fraction (0.1\%) data point.}
  \label{Fig/MethAr}
\end{figure}

High-statistics muon simulation studies were performed using the same independent Geant4 module originally developed for the estimation of in-situ cosmogenic backgrounds in the LEGEND-1000 pCDR. This module features a much more detailed representation of the LEGEND-1000 baseline design geometry, compared to the simple geometries used for the MCNP comparison study. Doping concentrations of methane in the outer LAr volume of 0.1, 1, 5, 10, and 20\% molar fraction were studied in the simulations. The rate of neutron capture on $^{76}$Ge in each simulation is normalized to the undoped case, as shown in Figure \ref{Fig/MethAr}. In the more aggressive doping cases of 10\% and 20\% molar fraction, reduction factors of about 3 and 5 were obtained in \gr~production, respectively, demonstrating the effectiveness of this hydrogen-rich liquid dopant. The neutron capture rates on argon atoms and on the hydrogen atoms in the methane molecules are also plotted separately. With more hydrogen in the LAr, neutrons are increasingly moderated and ultimately captured in these passive shielding materials instead of germanium. The decline in LAr neutron captures at higher methane concentrations is an effect of the increasing molar fraction of methane in the volume, which displaces some argon. The combined rate of neutron capture on the argon and hydrogen atoms continues to increase monotonically with methane concentrations. 

Methane doping within the LEGEND-1000 outer LAr volume would have both advantages and disadvantages. Practical implementation would be simple, as the outer LAr recirculation system and convection currents would gradually mix the dopant into the rest of the tank. Injecting a liquid dopant is expected to be inexpensive,
 with the main cost coming from sourcing the high-purity methane. Therefore, the doping can be made at a later stage of the LEGEND-1000 design, even post-assembly, allowing a final decision based on the outcome of LEGEND-200. Methane is a light molecule and exists in gaseous form at room temperature, facilitating the purification process and removal of heavy radio-impurities. On the other hand, even small concentrations of methane dopant would quench the scintillating properties of the pure atmLAr in the outer cryostat. 
 This could be mutually exclusive to proposed design changes which take advantage of scintillation from the outer LAr volume. Finally, there are controls on use of flammable materials in underground laboratories. P10 gas, which is a mixture of 10\% methane with 90\% argon, has been widely used in time projection chambers, \textit{e.g.} of the STAR experiment~\cite{Kotchenda2003}, and is considered a nonflammable substance, although it can burn in air under certain conditions~\cite{Haggerty2001}. However, with reasonable precautions and responsible implementation, heavy methane doping presents a viable risk reduction option if LEGEND-1000 is implemented at a candidate host site with significant cosmogenically-induced background. This study also provides a compelling motivation to consider other neutron moderators which are soluble in LAr.

\acknowledgments
This material is based upon work supported by the Particle Astrophysics Program and Nuclear Physics Program of the National Science Foundation through grant numbers PHY-1812356 and PHY-2111140. We gratefully acknowledge the support of the U.S.~Department of Energy Office of Science Graduate Student Research Program at the Los Alamos National Laboratory. We acknowledge research computation support from the University of South Dakota's Research Computing Group.

\end{document}